\documentstyle[aps,eclepsf]{revtex}
\begin{document}
\draft
\title{First Results from Dark Matter Search Experiment in the
Nokogiriyama Underground Cell}
\author{
\underline{W. Ootani} $^{\rm a}$\thanks{e-mail: \tt
ootani@postman.riken.go.jp},
M. Minowa $^{\rm bc}$, 
K. Miuchi $^{\rm b}$, 
Y. Inoue $^{\rm d}$,\\
T. Watanabe $^{\rm b}$,
M. Yoshida $^{\rm d}$,
Y. Ito $^{\rm b}$\thanks{Present address: KEK,
High Energy Accelerator Research Organization, 3-2-1 Midori-cho,
Tanashi-shi, Tokyo 188-8501, Japan},
Y. Ootuka $^{\rm e}$\thanks{Present address: Institute of Physics,
University of Tsukuba, Tsukuba, Ibaraki 305-0006, Japan}
}
\address{
$^{\rm a}$RIKEN, The Institute of Physical and Chemical Research, 2-1
Hirosawa, Wako, Saitama 351-0198, Japan\\
$^{\rm b}$Department of Physics, School of Science, University of
Tokyo, 7-3-1 Hongo, Bunkyo-ku, Tokyo 113-0033, Japan\\
$^{\rm c}$RESCEU, Research Center for the Early Universe, School of
Science, University of
Tokyo, 7-3-1 Hongo, Bunkyo-ku, Tokyo 113-0033, Japan\\
$^{\rm d}$International Center for Elementary Particle Physics, 7-3-1
Hongo, Bunkyo-ku, Tokyo 113-0033, Japan\\
$^{\rm e}$Cryogenic Center, University of Tokyo, 
2-11-16 Yayoi, Bunkyo-ku, Tokyo 113-0033, Japan}

\maketitle
\begin{abstract}
An experiment to search for hypothetical particle dark matter using
cryogenic thermal detector, or bolometer is ongoing. The bolometer
consists of eight pieces of 21\,g LiF absorbers and sensitive NTD
germanium thermistors attached to them and is installed in the
Nokogiriyama underground cell which is a shallow depth site ($\sim
15$\,m w.e.). We report on the results from the first running for
about ten days using this
arrayed bolometer system together with appropriate shieldings and muon
veto counters. From the obtained energy spectra the exclusion limits for the 
cross section of the elastic neutralino-proton scattering are derived
under commonly accepted astrophysical assumptions. The sensitivity for
the light neutralino with a mass below 5\,GeV is improved by this work. 
\end{abstract}
\vskip0.5cm

\section{Introduction}
There are a number of observational evidences to believe that a large fraction of the
matter in the Universe exists in the form of non-baryonic particle dark
matter. Supersymmetric neutralino is one of the most plausible
candidates for such exotic particle dark matter.
Various experimental efforts are being made aiming at detection of low energy nuclear
recoils caused by the elastic scatterings of
the neutralinos off nuclei\cite{dark matter search}. 

Conventional detectors like
semiconductor detectors or scintillators generally have a quenching
factor less than unity. Here the quenching factor is defined as
the ratio of the energy detection efficiency for a nuclear recoil to
that for an electron. On the other hand, because the bolometer
is sensitive to the whole energy deposited in the absorber the
quenching factor of the bolometer should be unity in
principle. Actually a quenching factor close to unity has been
measured by Milan group\cite{Alessandrello}. 

We have been developing bolometers with
lithium fluoride absorbers\cite{Tokyo}. Fluorine is
considered to have a large cross section for elastic scattering of
the axially-coupled neutralino off the nucleus compared with other
nuclei\cite{Ellis}. Recently we have
successfully constructed the bolometer array with a total mass of
168\,g and installed it in the Nokogiriyama underground cell with
a depth of 15\,m w. e.

In this paper we report on the first results from the experiment
performed in the Nokogiriyama underground cell using the bolometer array. 

\section{Experimental Set-up and Measurement}
\begin{figure}[p]
  \begin{center}
    \epsfile{file=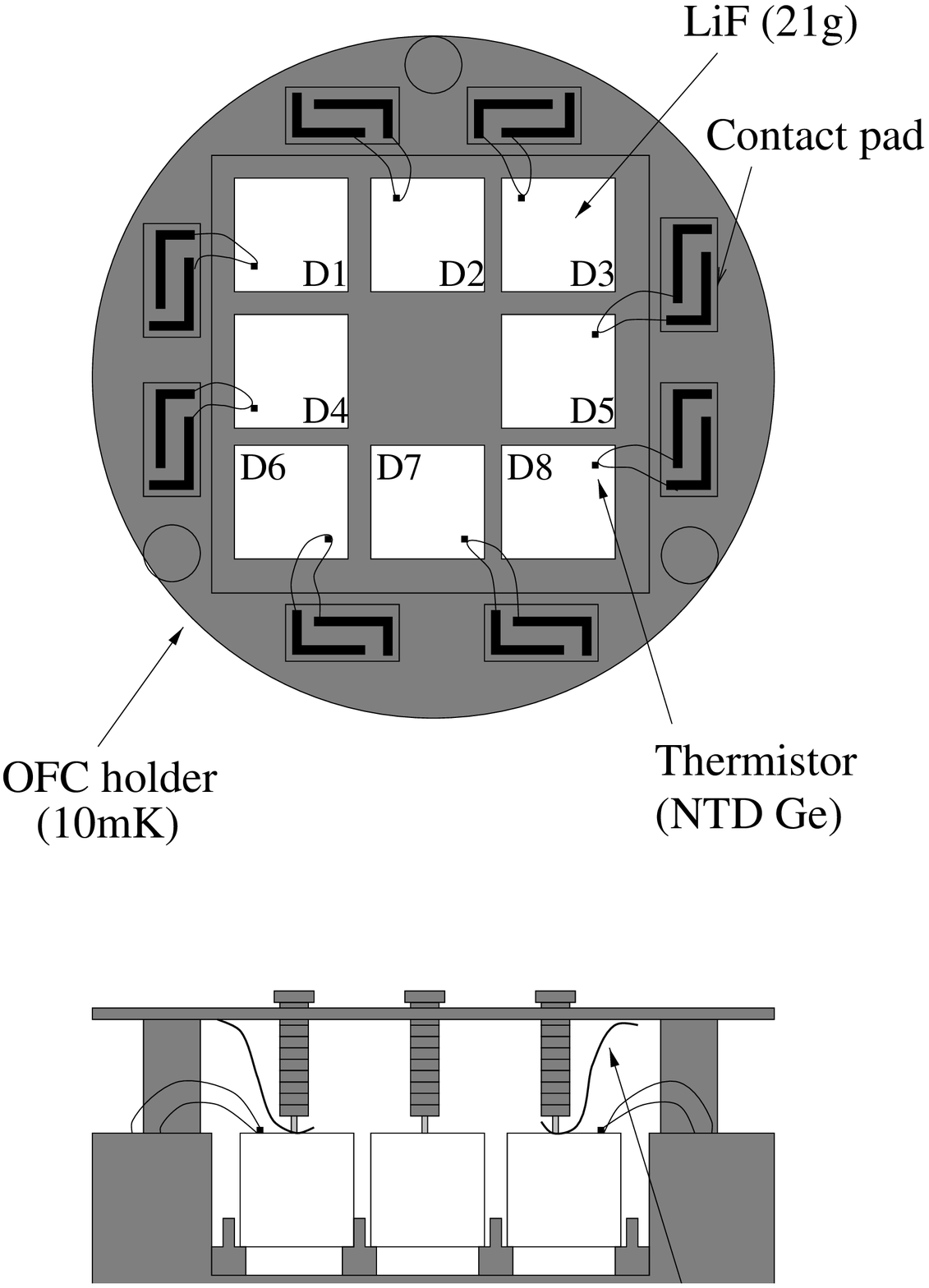,width=10cm}
    \caption{Schematic drawing of the lithium fluoride bolometer array.}
    \label{fig:multi}
  \end{center}
\end{figure}

The bolometer array used in this work contains eight 21\,g LiF
bolometers. The schematic drawing of the bolometer array
is shown in Fig.\,\ref{fig:multi}. The neutron transmutation doped
(NTD) germanium thermistors with the similar
temperature dependence of the resistance\cite{NTD} are attached to the
crystals. The thermistor senses a small temperature rise of the
absorber crystal caused by the neutralino-nucleus scattering. Each
crystal is placed on four copper posts and 
thermally insulated by the Kapton sheets. Moderate thermal anchoring of the
crystal to the copper holder with a temperature of 10\,mK is realized
by a oxygen free copper (OFC) ribbon. The lithium fluoride crystals are checked by
a low-background Ge spectrometer prior to the construction of the
bolometer. The concentration of radioactive contaminations is less
than 0.2\,ppb for U, 1\,ppb for Th, and 2\,ppm for K. The bolometer
array is mounted on a mixing chamber of a dilution refrigerator which
is mostly made of low-radioactivity materials radio-assayed in advance by low-background Ge spectrometer.

Each thermistor is biased through a 100\,M$\Omega$ load
resistor. The voltage change across the thermistor is fed into the
eight channel source follower circuit placed at the 4\,K
stage which include a low noise junction field effect transistor (J-FET),
Hitachi 2SK163. Since the J-FET does not work at this low
temperature, it is connected to a printed circuit board with thin stainless
steel tubes and manganin wires to be thermally isolated from the
circuit board with a temperature of 4\,K  and the temperature of the
FET is maintained above 100\,K by the heat produced by itself. 

The signal from the source follower circuit is in turn
amplified by an eight channel voltage amplifier placed just above
the refrigerator. 
The output of the voltage amplifier is fed into a double
pole low-pass filter with a cut-off frequency of 226\,Hz and in turn into the 16-bit waveform digitizer to record the pulse shape of
the signal for off-line analysis.

The passive radiation shielding consists of
10\,cm-thick oxygen free high conductivity copper layer, 15\,cm-thick
lead layer, 1\,g cm$^{-2}$-thick boric acid layer and 20\,cm-thick
polyethylene layer. The latter two layers act as a neutron shield. In
order to avoid muon-induced background we employ a veto system
which consists of 2\,cm-thick plastic scintillators.  

The constructed detector system is installed in the Nokogiriyama
underground cell which is located about 100\,km south from Tokyo and
relatively easy to access. The depth of an overburden of sand is inferred to be
about 15\,m w.e. In this work six
bolometers of the  bolometer array are
used and energy spectrum are measured for about ten days. Two
bolometers have some problems in the
cooling procedure. 

\begin{figure}[p]
  \begin{center}
    \epsfile{file=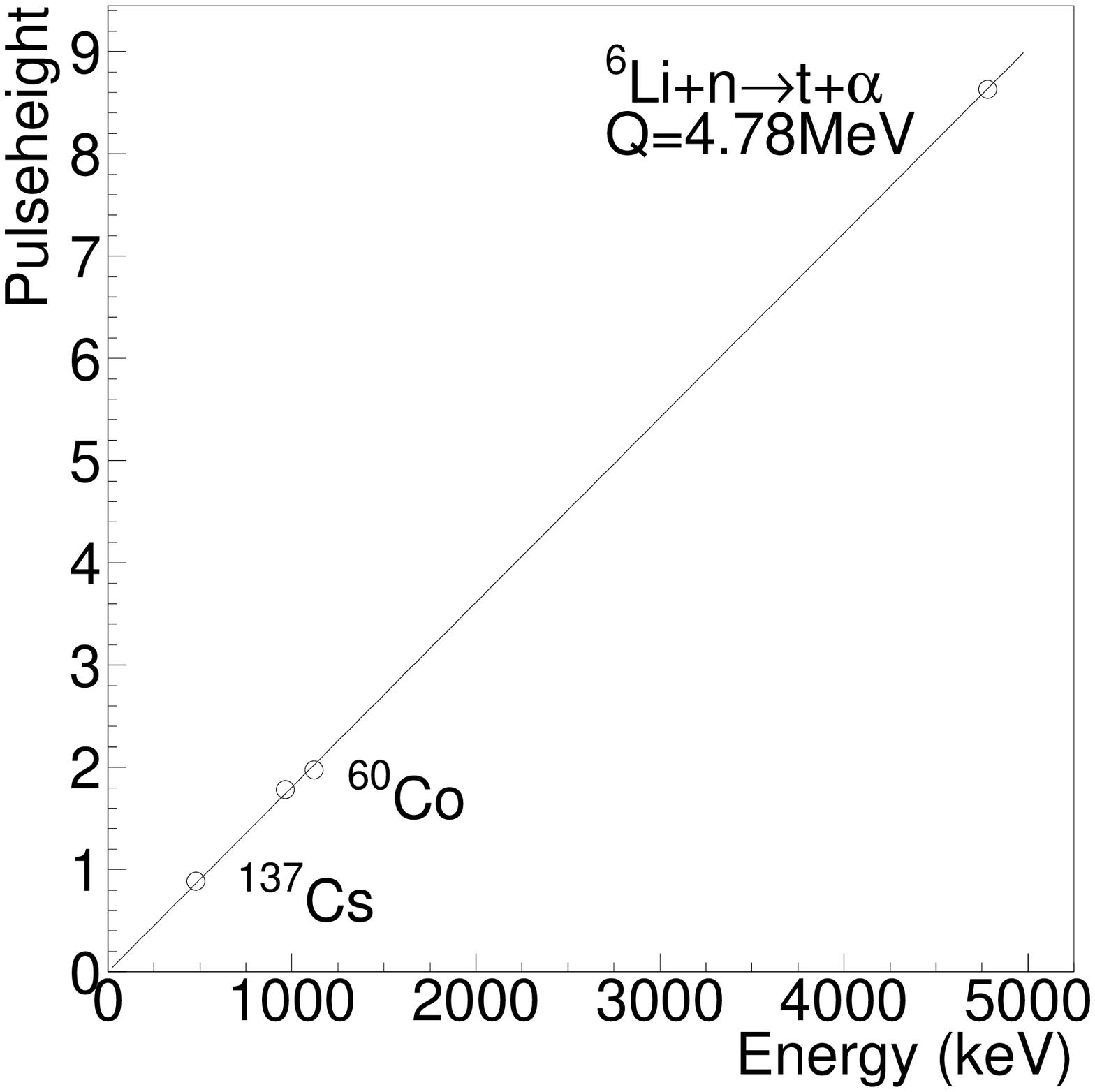,width=14cm}
    \caption{Linearity obtained for one of six bolometers.}
    \label{fig:calibration}
  \end{center}
\end{figure}

Since the detector is enclosed in a cryogenic vacuum can during the 
measurements it is impossible to place the gamma-ray source close to
the detector for energy calibration. The energy calibration during the
measurements is, therefore, performed by 662\,keV gamma-rays from a
$^{137}$Cs source and 1333\,keV and 1173\,keV gamma-rays from a
$^{60}$Co source placed outside a helium dewar of the dilution refrigerator
and inside the radiation shieldings.
Furthermore, the sharp peak at 4.78\,MeV due to the neutron capture
reaction of $^6$Li observed in the background spectrum is also used for 
pthe energy calibration. Fig.\,\ref{fig:calibration} shows one
of the obtained energy calibration plots. Linearities of the six
bolometers up to 5\,MeV are recognized.  It must be noted
that linearity down to 60\,keV gamma-ray is confirmed prior to this
measurement using gamma-ray from $^{241}$Am source set inside the cryostat.

\section{Energy Spectra and Dark Matter Limits}
\begin{figure}[p]
  \begin{center}
    \epsfile{file=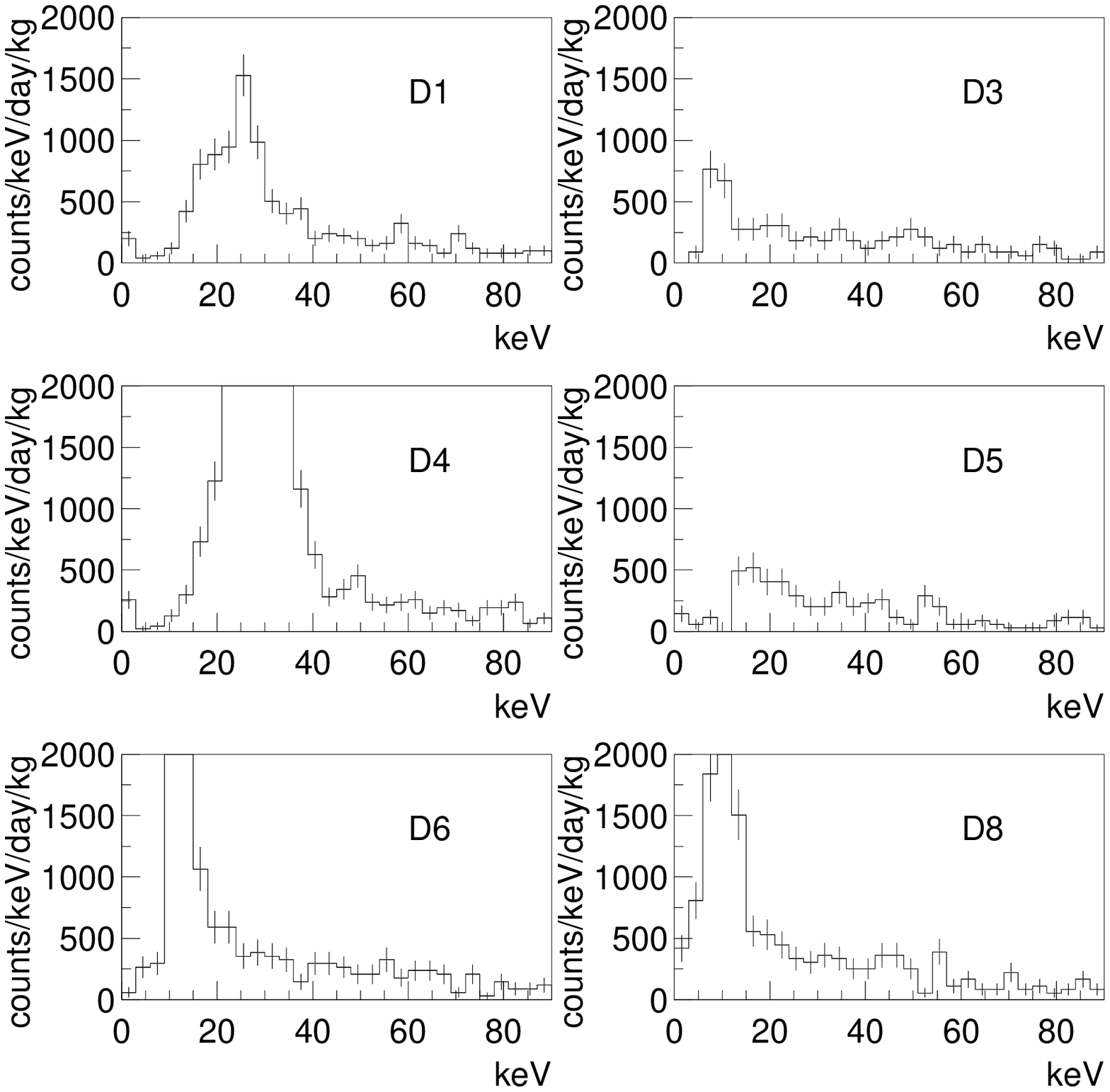,width=16cm}
    \caption{Energy spectra obtained by six LiF bolometers.}
    \label{fig:spectrum}
  \end{center}
\end{figure}

Fig.\,\ref{fig:spectrum} shows the energy spectra obtained by
the six bolometers during ten days. The
bump in the low energy region is considered to be due to microphonics
caused by a helium liquefier which recondenses evaporated 
helium gas from the dewar. While the similar spectra are obtained for 
four bolometers (D3, D5, D6, and D8), the spectra for the other two
bolometers (D1 and D4) are affected
by microphonics below 30 to 40\,keV because of their low detector gains.

Comparing the measured energy spectrum with the expected recoil spectrum, the
exclusion limits for the cross section for elastic neutralino
scattering off the nucleus can be extracted. The calculation is
performed in the same manner used in Ref.\,\cite{smith}. The theoretical recoil
spectrum is calculated 
assuming a Maxwellian dark matter velocity distribution with rms
velocity of 230\,km/s, and then folded with the measured energy
resolution and the nuclear form factor. 
We also assume the local halo density of the neutralino to be
0.3\,GeV/cm$^3$. The spin factors calculated assuming an odd group
model as a nuclear shell model are 0.75 for $^{19}$F and 0.417 for
$^{7}$Li\cite{Ellis}. Since the detector responses for the six
bolometers are not the same, the upper limit of the cross
section is evaluated independently from the
spectrum of each detector. For a given neutralino mass the lowest
value of the cross section is taken as a combined limit from the
results of the six detectors.

\begin{figure}[p]
  \begin{center}
    \epsfile{file=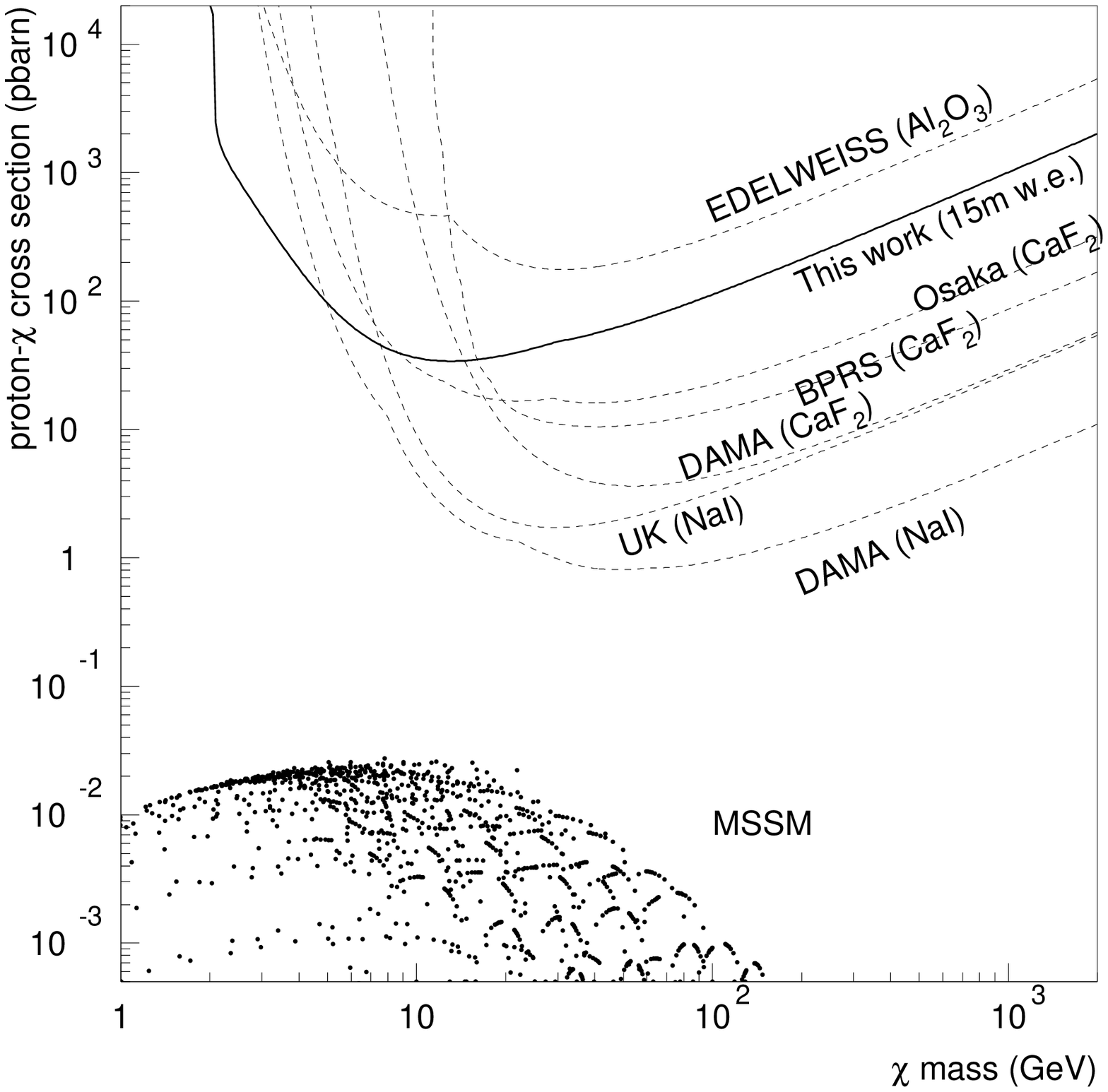,width=16cm}
    \caption{Exclusion limit obtained from this experiment for
      spin-dependent interaction as a function of neutralino mass.}
    \label{fig:limit}
  \end{center}
\end{figure}

The calculated exclusion limits in case of the
spin-dependent interaction are given in Fig.\,\ref{fig:limit}. For comparison the exclusion limits derived from the data
in the other experiments at deep underground
sites\cite{EDELWEISS,BPRS,DAMA,smith,OSAKA} and the scatter plots
predicted in the minimal supersymmetric theories are also shown.  
Although the other experiments except for Osaka experiment are
performed at deep underground laboratories, our experiment gives comparable limits for the light neutralino. This
owes to the large cross section for the spin-dependent interaction of
$^{19}$F and the low energy threshold of the bolometer. 
The sensitivity for neutralinos with a mass below 5\,GeV is
improved by this work. 
\section{Prospects}
Compton scattered gamma-rays from the aperture of the shielding and 
gamma-rays produced through the interaction of cosmic ray muon within the 
shielding materials are considered  major background sources.  
In order to reduce the muon-correlated background, the veto
efficiency must be improved. The present incompleteness of the veto is
due to the penetrations for vacuum tubes and the tube of the helium liquefier. 
Increasing of the coverage of the plastic scintillator will improve
the veto efficiency up to 98\%. Against the Compton scattered gamma-ray
background, internal lead shielding with a thickness of 20\,mm
surrounding the lithium fluoride
bolometer array will be installed. The shielding is made of over 200 year old low-activity lead with a concentration of $^{210}$Pb of
less than 0.05\,pCi/g. The Compton scattered
gamma-rays can be reduced by two orders of magnitude by this internal
shielding. Since lead fluorescence X-rays are produced mainly by the
muon interaction, their contribution can be ignored if muons are sufficiently vetoed. If
these improvements are realized, the sensitivity of
this experiment will be improved by more than an order of magnitude
even at this shallow depth. 

The detector system will be installed
in a underground facility with a sufficient depth where cosmic muon
induced background is expected to be negligible. The long-term
measurements in a deep underground site will bring the sensitivity to
the spin-dependent interaction below the level predicted by
the supersymmetric theory.
\section*{Acknowledgments}
We would like to thank Prof. Komura for providing us with the
low-radioactivity old 
lead. This research is supported by the Grant-in-Aid for COE Research by the 
Japanese Ministry of Education, Science, Sports and Culture. W.O. are
grateful to Special Postdoctoral Researchers Program for support of
this research.

\newpage

\end{document}